\documentstyle[12pt,epsf,epsfig,wrapfig]{article}
\newcommand{\ba}{\begin{eqnarray}}\newcommand{\ea}{\end{eqnarray}}

\newcommand{\be}{\begin{equation}}
\newcommand{\ee}{\end{equation}}

\def\xb{\overline{x}}

\def\gev{\,{\rm GeV}}
\begin{document}

\begin{center}
{\bfseries ELECTROPRODUCTION OF LIGHT VECTOR MESONS}

\vskip 5mm S.V.Goloskokov $^{\dag}$

\vskip 5mm {\small {\it Bogoliubov Laboratory of Theoretical
Physics, Joint Institute
for Nuclear Research,\\ Dubna 141980, Moscow region, Russia}}\\
$\dag$ {\it E-mail:goloskkv@theor.jinr.ru }
\end{center}

\vskip 5mm
\begin{abstract}
An analysis of light vector meson photoproduction at small Bjorken
$x \leq 0.2$ is done on the basis of the generalized parton
distributions (GPDs). Our results on the cross section and spin
density matrix elements (SDME) are in good agreement with
experiments.
\end{abstract}

\vskip 8mm

This report is devoted to the study of the vector meson
leptoproduction  at Bjorken $x \leq 0.2$  based on our results
\cite{gk05,gk06,gk07t}. At large photon virtualities   the
amplitude for longitudinally polarized virtual photons and vector
meson (LL amplitude) factorizes \cite{fact} into a hard meson
photoproduction off partons and GPDs. Unfortunately, in the
collinear approximation the LL cross section exceeds the data by
an order of magnitude \cite{mpw}. Moreover, in this approximation
the amplitude for transversally polarized photons (TT amplitude)
exhibits infrared singularities \cite{mp}, which signals the
factorization breakdown.

In this report, we  discuss the spin effects in the vector meson
leptoproduction. Our calculations \cite{gk06,gk07t} are  based on
the modified perturbative approach (MPA) \cite{sterman} which
includes the quark transverse degrees of freedom accompanied by
Sudakov suppressions. The contribution from the end-point region
to the LL amplitude is suppressed in our model  and the cross
section is close to the experiment. The TT amplitudes  can be
calculated in the model because the transverse quark momentum
regularizes the
 singularities.  Within the MPA we calculate the  cross
sections and the spin observables in the energy range $5
\mbox{GeV} < W < 90 \mbox{GeV}$. Our results on the cross section
and SDME are in good agreement with  experiments \cite{h1, zeus,
hermes, compass}.

The model is based on the handbag approach where the $\gamma^{\ast
} p\rightarrow V p$ amplitude  factorizes into hard partonic
subprocess and GPDs. In the region of small $x \leq 0.01$  gluons
give the dominant contribution \cite{gk05}. At larger $x \sim
0.2$, in addition to the gluon GPD the inclusion of quark
contribution is important \cite{gk06,gk07t}. For small $t$ the
amplitude of the vector meson production off the proton with
positive helicity reads as a convolution of the partonic
subprocess
 ${\cal H}^V$ and GPDs $H^i\,(\widetilde{H}^i)$
\begin{eqnarray}\label{amptt-nf-ji}
  {\cal M}^V_{\mu'+,\mu +} &=& \frac{e}{2}\, {\cal
  C}^{V}\, \sum_{\lambda}
         \int d\xb
        {\cal H}^{Vi}_{\mu'\lambda,\mu \lambda}
                                   H^i(\xb,\xi,t) ,
\end{eqnarray}
where  $i$ denotes the gluon and quark contribution,
 $\mu$ ($\mu'$) is the helicity of the photon (meson), $\xb$
 is the momentum fraction of the
parton with helicity $\lambda$, and the skewness $\xi$ is related
to Bjorken-$x$ by $\xi\simeq x/2$. The flavor factors are
$C^{\rho}=1/\sqrt{2}$ and ${ C}^{\phi}=-1/3$. In the analysis of
the cross section  at small $x$ the main contribution is
determined by  the unpolarized GPDs $H^i$.

The $k$- dependent wave function \cite{koerner} that contains the
leading and higher twist terms   describing the longitudinally and
transversally polarized vector meson is used to calculate the
partonic subprocess ${\cal H}$ in (\ref{amptt-nf-ji}); ${\cal H}$
is estimated within the MPA \cite{sterman} where we keep the
$k^2_\perp$ terms in the denominators of the amplitudes and in the
numerator of the TT amplitude. The gluonic corrections are treated
in the form of the Sudakov factors which additionally suppress the
end-point integration regions.

The GPDs   are modeled using the double distribution
\begin{equation}
  H_i(\xb,\xi,t) =  \int_{-1}
     ^{1}\, d\beta \int_{-1+|\beta|}
     ^{1-|\beta|}\, d\alpha \delta(\beta+ \xi \, \alpha - \xb)
\, f_i(\beta,\alpha,t).
\end{equation}
Here the double distribution function $f_i(\beta,\alpha,t)$ is
connected with the corresponding parton distributions (PDFs) which
are taken from the CTEQ6M results \cite{CTEQ}. The simple Regge
ansatz is used to consider $t$ dependencies of PDFs. For details
see \cite{gk06,gk07t}.

\begin{figure}[h!]
\begin{center}
\begin{tabular}{cc}
\mbox{\epsfig{figure=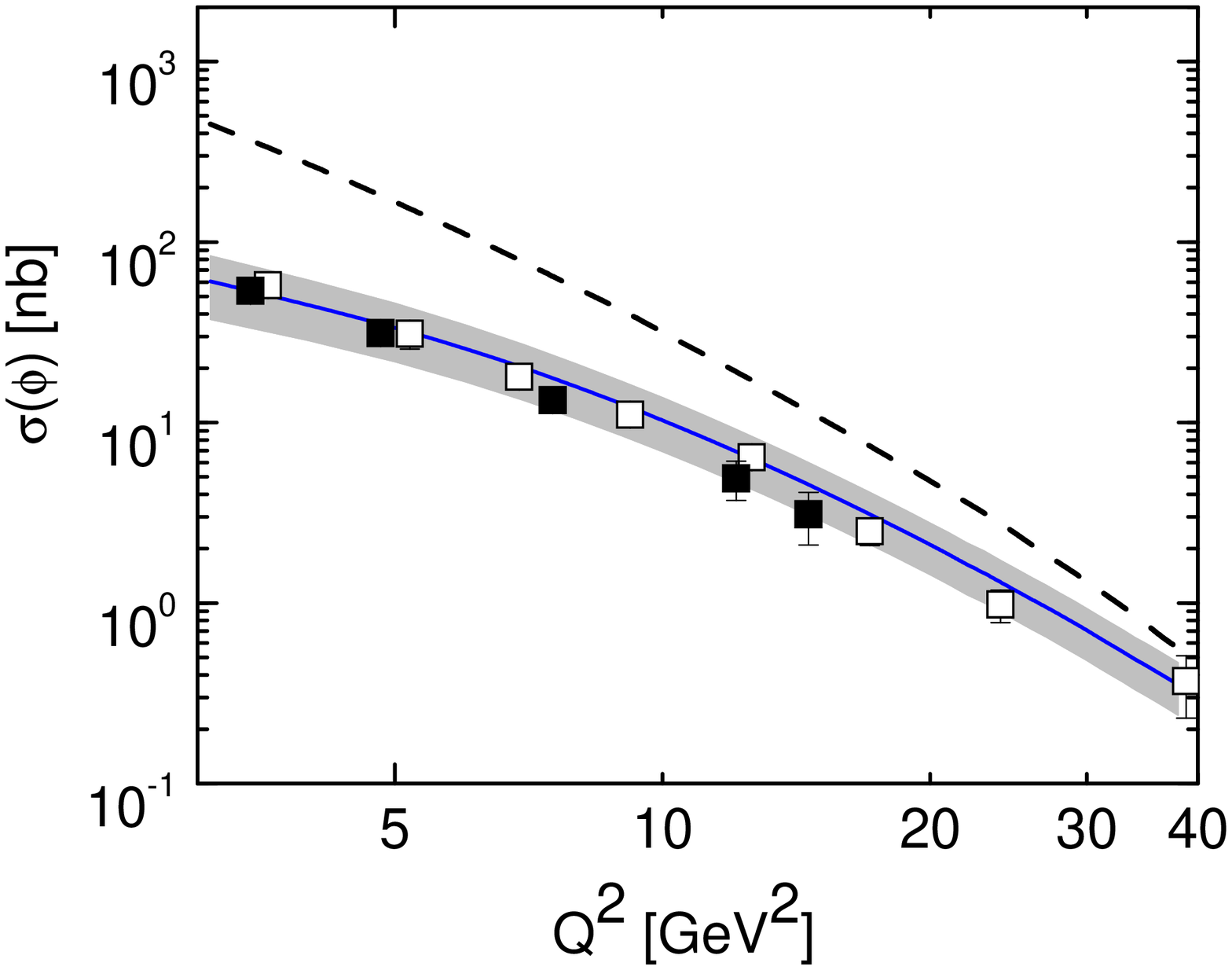,width=7.7cm,height=6cm}}&
\mbox{\epsfig{figure=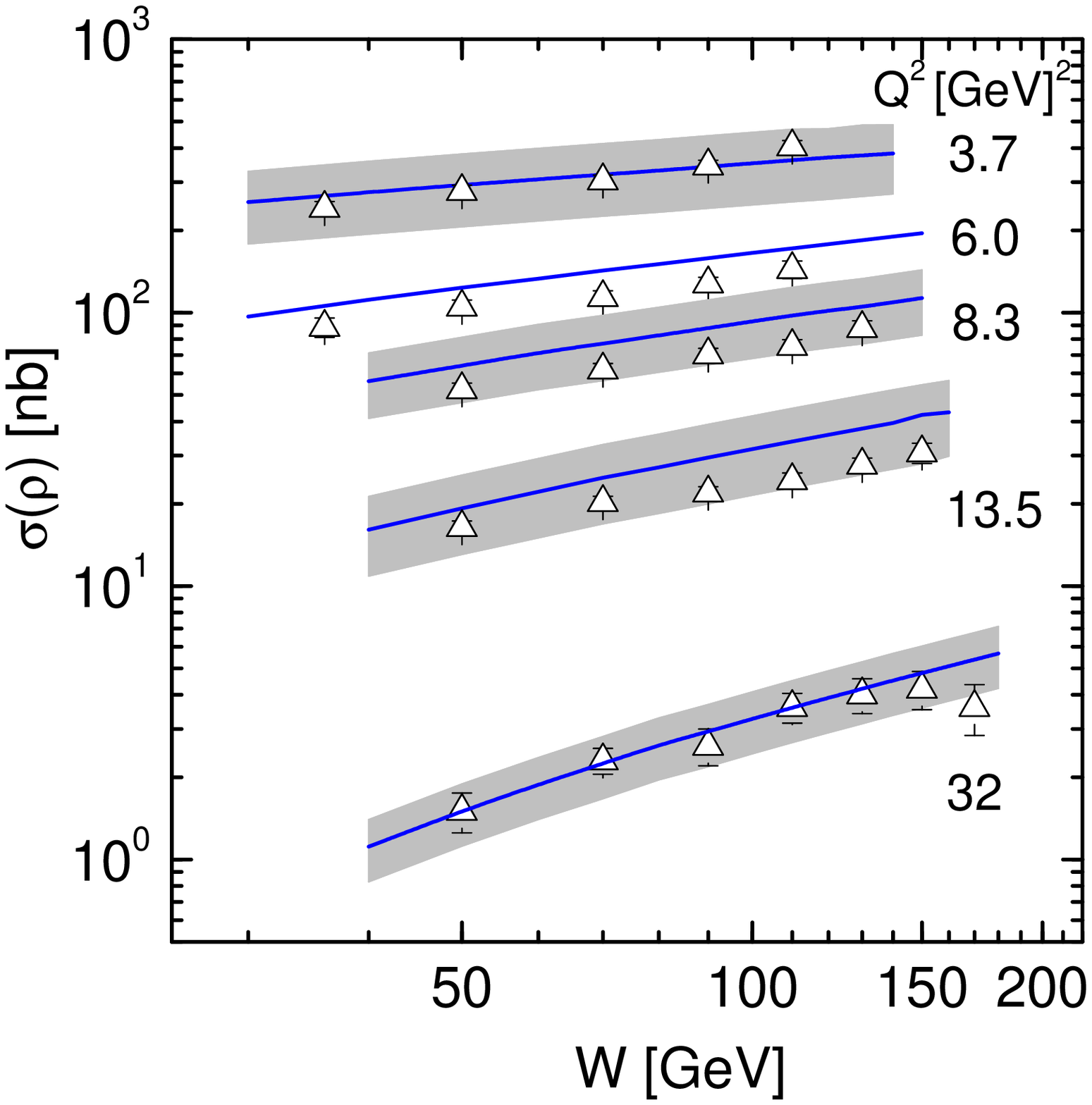,width=7.5cm,height=6.2cm}}
\end{tabular}
\end{center}
{\small{\bf Figure 1:} Left: The  cross sections of $\phi$
production at $W=75 \mbox{GeV}$ with error band from CTEQ6 PDFs
uncertainties. Data are from H1 \cite{h1} -solid symbols and ZEUS
\cite{zeus} -open symbols. Dashed line- LO result. Right: The
cross sections of $\rho$ production   via $W$ at different $Q^2$.}
\end{figure}

The cross section for the $\gamma^* p \to \phi p$
 production integrated over $t$  is shown
in Fig.1 (full line). Good agreement with DESY experiments
\cite{h1,zeus} is  observed.  The shared bands in the figures
reflect uncertainties of our results caused by the errors in the
CTEQ6 PDFs. The leading twist results  are also presented in Fig.
1. The $k_\perp^2/Q^2$ corrections in the hard amplitude decrease
the cross section by a factor of about 10 at $Q^2 \sim
3\mbox{GeV}^2$.

Our results reproduce well the energy dependence of the $\rho$
cross section \cite{zeus} as shown in Fig.1. The cross section at
HERA energies is dominated by the gluon and sea quark
contributions.

\begin{figure}[h!]
\begin{center}
\begin{tabular}{ccc}
\mbox{\epsfig{figure=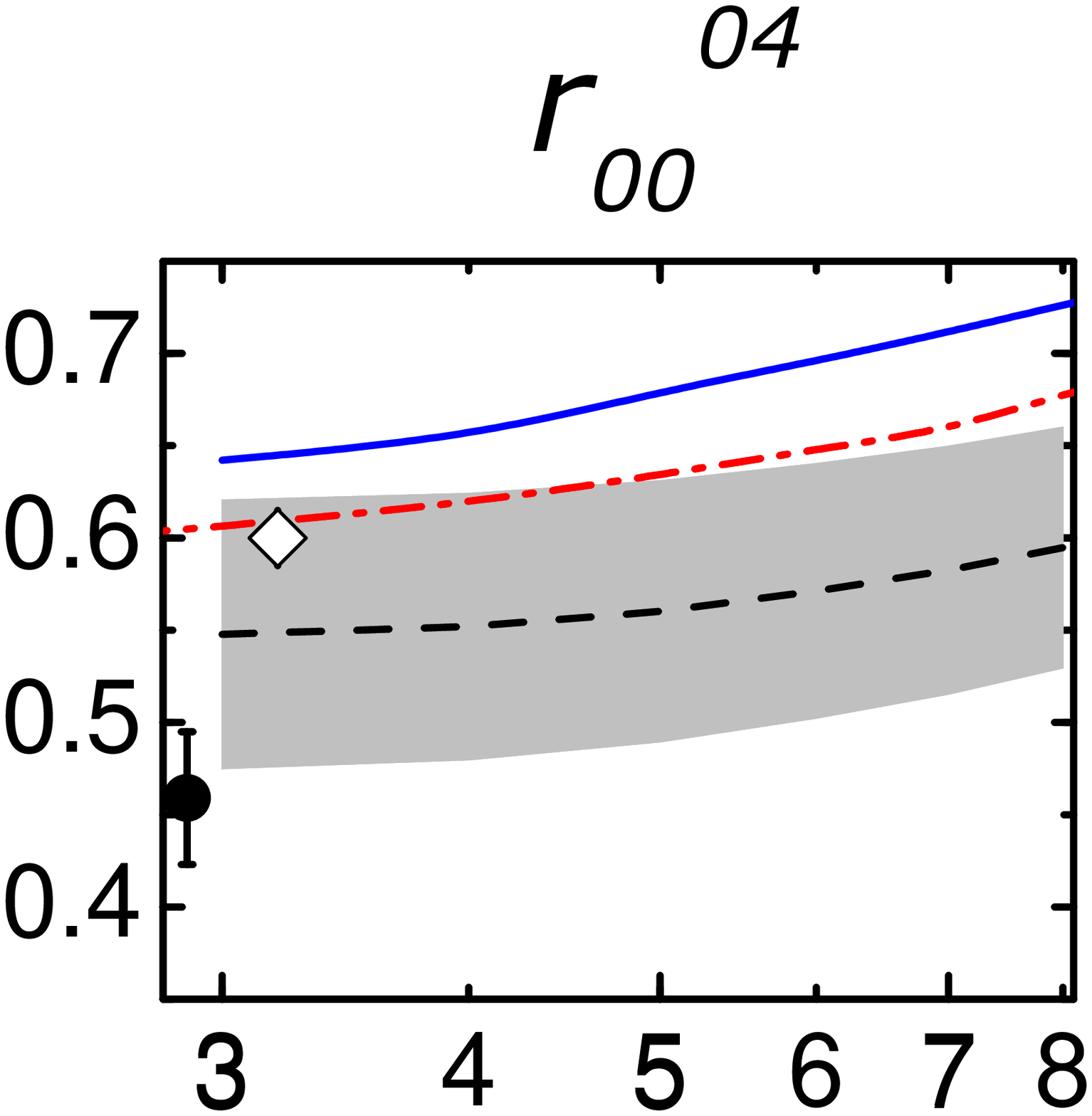,width=4.6cm,height=4.3cm}}&
\mbox{\epsfig{figure=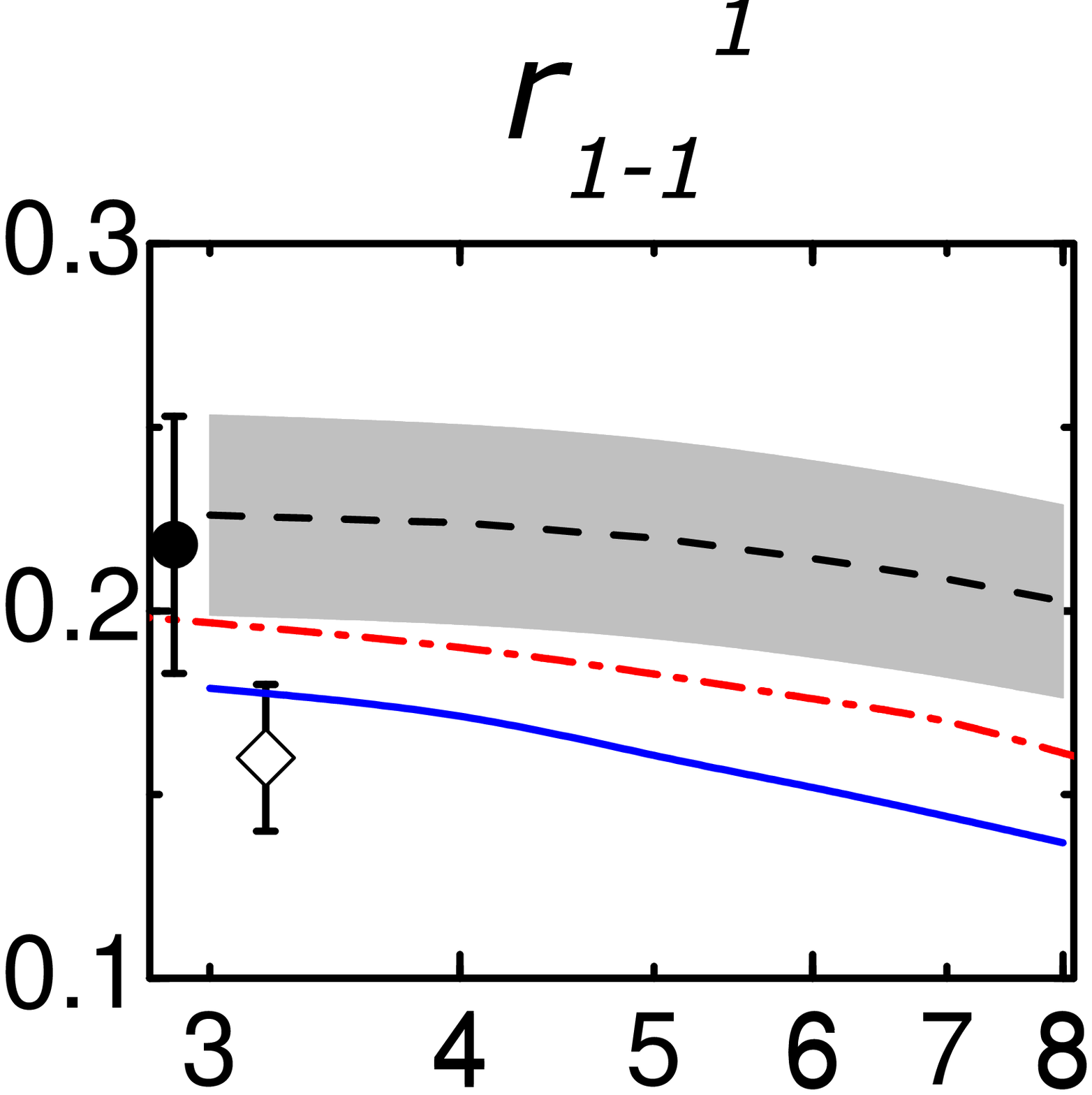,width=4.6cm,height=4.3cm}}&
\mbox{\epsfig{figure=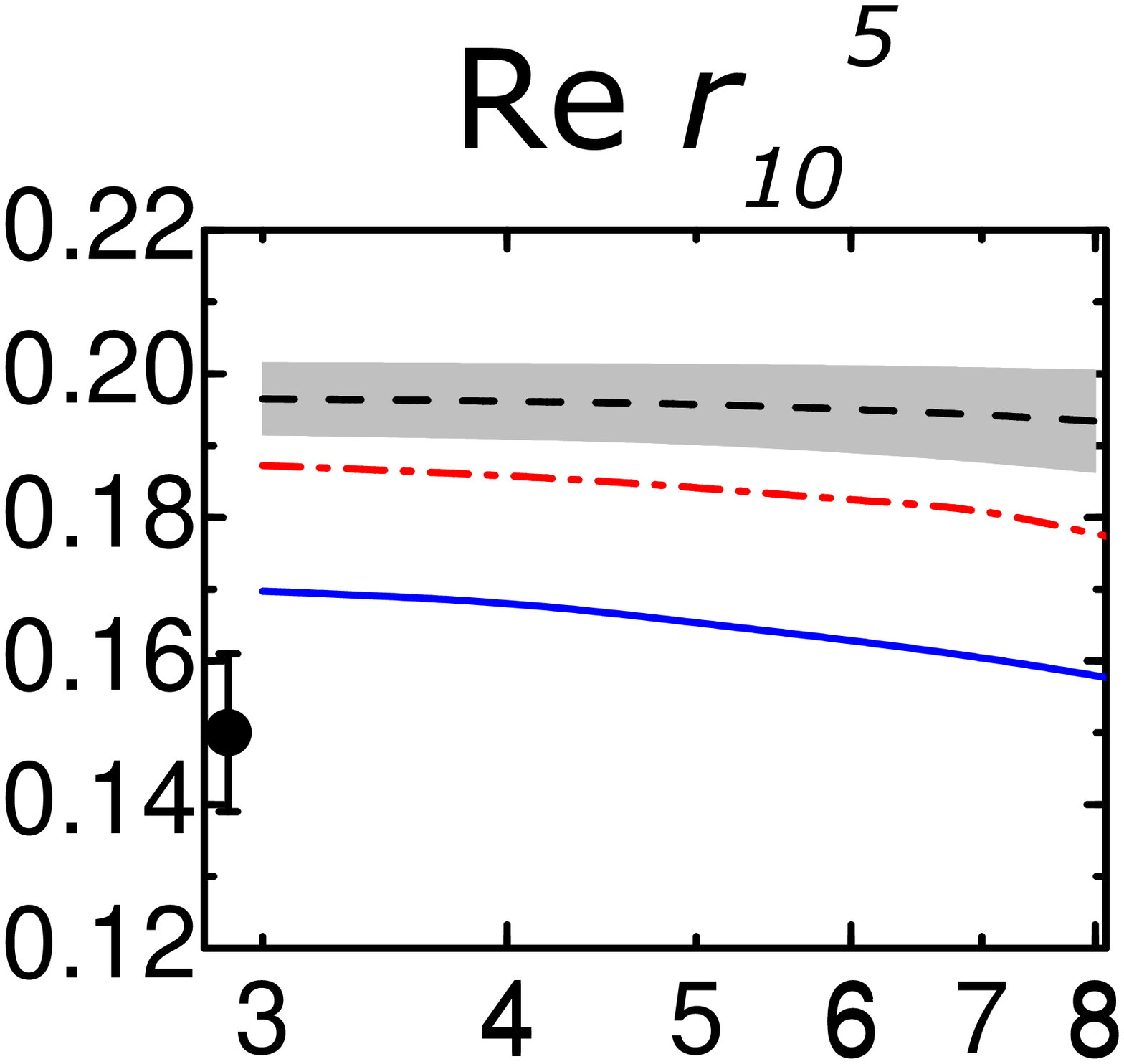,width=4.6cm,height=4.3cm}}\\
\end{tabular}
\phantom{aa}\vspace{-3mm}
\end{center}
{\small{\bf Figure 2:} The $Q^2$ dependence of SDME on the $\rho$
production and $W=75(10,5)\gev$- solid(dash-dotted, dashed) line.
Preliminary data are taken from HERMES \cite{hermes} (solid
circles) and COMPASS \cite{compass} (diamonds).}
\end{figure}

The model describes properly spin effects determined by the TT
transition amplitude. Our results for the ratio of the
longitudinal and transverse cross sections and SDME in the energy
range $5\mbox{GeV}< W < 75\mbox{GeV}$ can be found in
\cite{gk07t}. In Fig.2, we present the SDME on the $\rho$
production at  $W=5,10,75 \mbox{GeV}$. At HERMES energy $W =5
\mbox{GeV}$ the valence quark contribution to the amplitudes is
essential. At COMPASS $W =10 \mbox{GeV}$ quark effects are not so
large and they are negligible at HERA $W =75 \mbox{GeV}$. This is
the main reason of the energy dependencies of SDME shown in Fig.2.
A similar energy dependence is observed experimentally.

\begin{wrapfigure}[16]{hr!}{6.1cm}
\begin{center}
\phantom{.}\vspace{-1cm}
\mbox{\epsfig{figure=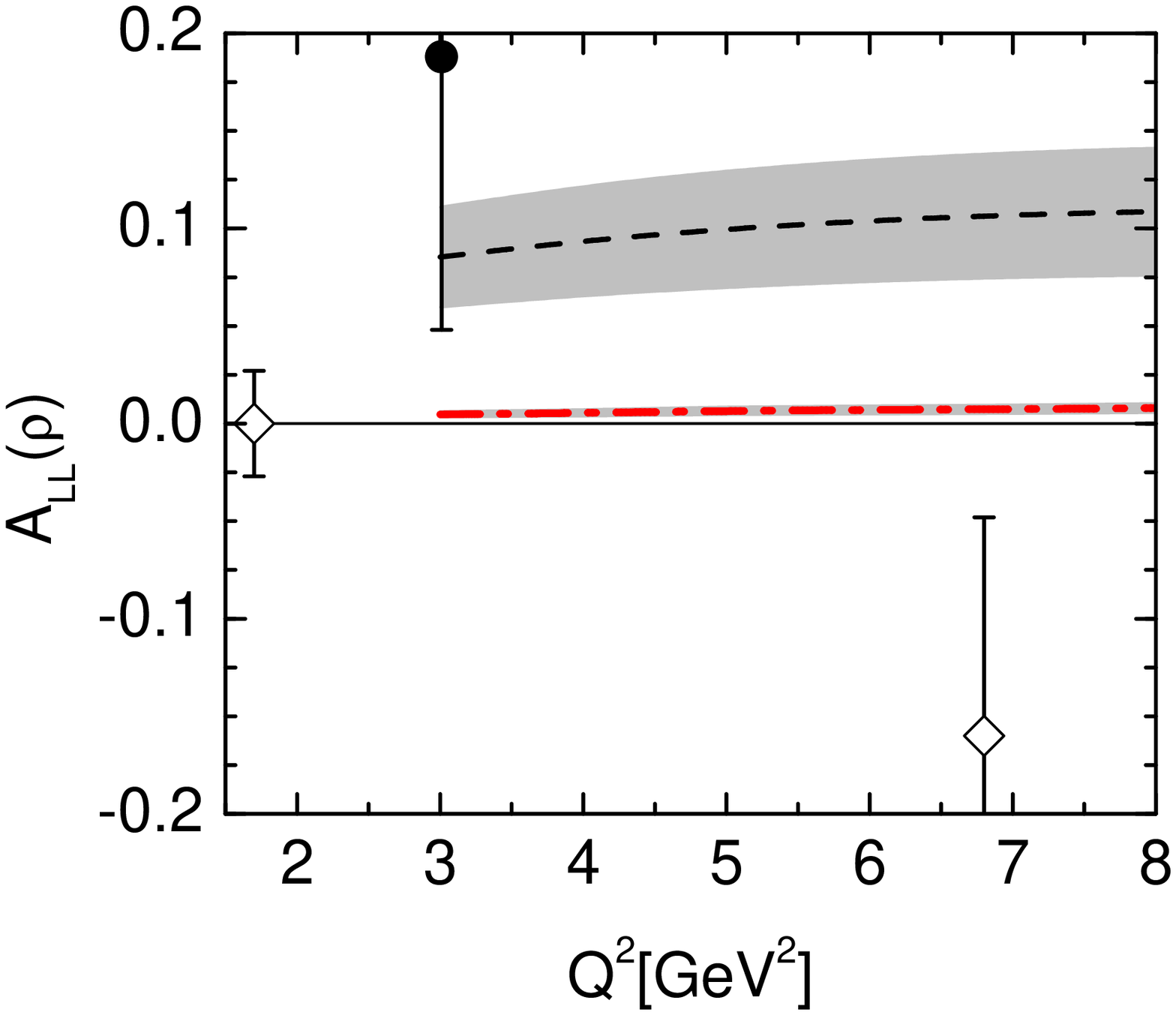,width=6cm,height=4.8cm}}
\end{center}
\vspace{-4mm} {\small{\bf Figure 3:} The $A_{LL}$ asymmetry for
the $\rho$ production at $W=5\,\gev$ (dashed line) and
  $W=10\,\gev$ (dashed-dotted line).}
\end{wrapfigure}

The $A_{LL}$ asymmetry for a longitudinally polarized beam and
target is sensitive to the polarized GPD.  The leading term in
$A_{LL}$ asymmetry integrated over the azimuthal angle is
determined through the interference between the $H$ and
$\widetilde{H}$ distributions. In Fig. 3, we show our results for
the $\rho$ production at $W=5 \mbox{GeV}$ and $W=10 \mbox{GeV}$.
At HERMES energies the valence quark contribution generates large
asymmetry of the order of 0.1 which is compatible with the
experimental results \cite{hermes}. At COMPASS \cite{compass}, the
valence quark contribution is small and asymmetry close to zero is
predicted. Note that we observe an essential cancellation of the
gluon and sea quark contributions. This leads to small $A_{LL}$
asymmetry for the $\phi$ production.

{\bf In summary}: Light vector meson electroproduction at small
$x$ was analyzed here within the  GPD approach. The  partonic
subprocesses have been calculated  using the MPA with the wave
function  dependent on the transverse quark momentum. The higher
order $k_\perp^2/Q^2$ corrections which are considered in the
propagators of the partonic subprocess  decrease the cross section
by a factor of about 10 at $Q^2 \sim 3\mbox{GeV}^2$. The same
higher order effects in the denominators of the hard subprocess
regularize the singularities in the TT amplitude. This gives a
possibility to calculate the TT amplitude and study spin effects
in the vector meson production in our  model.

In our previous calculations \cite{gk05} we analysed the low $x
\leq 0.01$ region where the gluon contribution has a predominant
role. In this report, we extend our analysis to  $x \sim 0.2$
\cite{gk06, gk07t}. In the moderate $x$ region we consider gluon,
sea and valence quark GPDs.  The GPDs are modeled via double
distribution based on the CTEQ6M  distributions. In the model we
find a good description of the cross section and the spin
observables from HERMES to HERA energies \cite{gk06}. It is found
that the gluon and sea contributions control the amplitude
behaviour at energies $W \geq 10 \mbox{GeV}$. Valence quarks are
essential only at HERMES energies, where their contribution to the
$\rho(\omega)$ cross section is about 40(65\%). This shows that
the $\omega$ production at low energies is much more sensitive to
valence quarks than $\rho$ production.

The model describes well the  ratio of the longitudinal and
transverse cross sections and SDME in the energy range
$5\mbox{GeV}< W < 75\mbox{GeV}$  \cite{gk07t}. We predict large
$A_{LL}$ asymmetry at  HERMES energies determined by the valence
quark contribution which is compatible with experiment. At COMPASS
the $A_{LL}$  asymmetry is small, about zero. Our first results on
SDME for transversally polarized target  and $A_{UT}$ asymmetry
cans be found in \cite{gk07t}.

Thus, we can conclude that the vector meson photoproduction at
small $x$ is a good tool to probe the   GPDs. Study   of SDME
gives important information on the structure of different helicity
amplitudes in the vector meson production.

 This work is supported  in part by the Russian Foundation for
Basic Research, Grant 06-02-16215  and by the Heisenberg-Landau
program.

\end{document}